\documentclass[%
 reprint,
superscriptaddress,
 amsmath,amssymb,
 aps,
]{revtex4-2}
\usepackage{xcolor}

\usepackage{graphicx}
\usepackage{dcolumn}
\usepackage{bm}

\begin{document}

\preprint{APS/123-QED}

\title{Characteristic exciton energy scales in antiferromagnetic NiPS$_3$}

\author{Jacob A. Warshauer}
\email{warsh@bu.edu}
\author{Huyongqing Chen}
\affiliation{Department of Physics, Boston University, 590 Commonwealth Avenue, Boston, MA 02215, USA}
\author{Qishuo Tan}
\affiliation{Department of Chemistry, Boston University, 590 Commonwealth Avenue, Boston, MA 02215, USA}
\author{Jing Tang}
\affiliation{Department of Chemistry, Boston University, 590 Commonwealth Avenue, Boston, MA 02215, USA}
\author{Xi Ling}
\affiliation{Department of Chemistry, Boston University, 590 Commonwealth Avenue, Boston, MA 02215, USA}
\affiliation{Division of Materials Science and Engineering, Boston University, 590 Commonwealth Avenue, Boston, MA 02215, USA}
\affiliation{Photonics Center, Boston University, 8 Saint Mary's St., Boston, MA 02215, USA}
\author{Wanzheng Hu}
\email{wanzheng@bu.edu}
\affiliation{Department of Physics, Boston University, 590 Commonwealth Avenue, Boston, MA 02215, USA}
\affiliation{Division of Materials Science and Engineering, Boston University, 590 Commonwealth Avenue, Boston, MA 02215, USA}
\affiliation{Photonics Center, Boston University, 8 Saint Mary's St., Boston, MA 02215, USA}

\begin{abstract}
    Two-dimensional antiferromagnets are promising materials for spintronics. The van der Waals antiferromagnet NiPS$_3$ has attracted extensive interest due to its ultra-narrow exciton feature which is closely linked with the magnetic ordering. Here, we use time-resolved terahertz spectroscopy to investigate photo-excited carriers in NiPS$_3$. We identify the onset of interband transitions and estimate the exciton dissociation energy from the excitation wavelength and fluence dependence of the transient spectral weight. Our results provide key insights to quantify the exciton characteristics and validate the band structure for NiPS$_3$. 
\end{abstract}

\maketitle
The van der Waals antiferromagnets have received significant attention for their strong potential to be a building block in two-dimensional electronic and spintronic devices \cite{Sierra2021, Mak2016} and as nonlinear photonics devices \cite{Yue2024}. In this material family, NiPS$_3$ is a particularly interesting case due to its ultra-narrow linewidth exciton \cite{Kang2020, Wang2021, Hwangbo2021}. Although the origin of this exciton is still under debate \cite{Kang2020, Jana2023, Klaproth2023, He2024, Kim2023, Li2024, Hamad2024}, the consensus is that this exciton couples to magnetism \cite{Hwangbo2021, Jana2023, Dirnberger2022, Wang2024, Song2024} and is clearly different from the excitons in conventional semiconductors. 

Characterization of the bandgap and the exciton dissociation energy of NiPS$_3$ is crucial for understanding the unconventional excitonic physics in this material. In general, bandgap information can be obtained by equilibrium spectroscopic measurement. In reality, the bandgap may be obscured due to the complex electronic structure in strongly correlated materials, where many excited-states of varied origin may overlap as a result of their close proximity in energy \cite{Kim2018, Tan2022, Wang2022}. To date, the existing literature reports diverse values of the bandgap for NiPS$_3$ \cite{Brec1979, Ramesh2016, Lane2020, Wang2021, Ho2021, Klaproth2023, Cao2024}. As for the exciton dissociation energy, a large exciton binding energy at the eV level has been predicted in this material family at the monolayer limit \cite{Rybak2024}. Experimentally, the narrow linewidth exciton persists down to the trilayer, but no exciton feature is seen in the monolayer \cite{Hwangbo2021}. Furthermore, the exciton dissociation energy for bulk antiferromagnetic NiPS$_3$ remains unknown. Therefore, probing the exciton band structure in NiPS$_3$ is in great demand. 

\begin{figure*}
\includegraphics[scale=1]{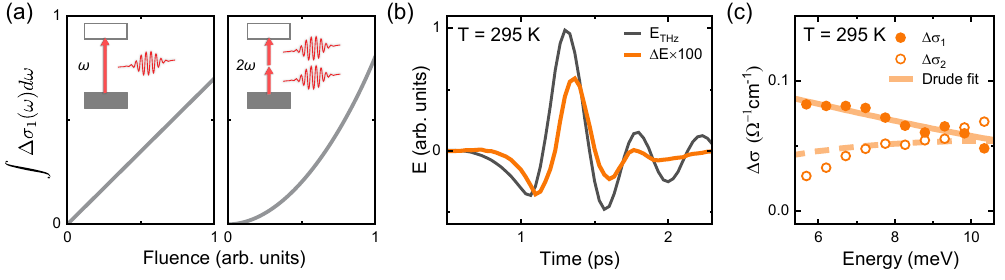}
\caption{A general approach to identify the bandgap and a typical pump-probe response for NiPS$_3$. (a) Excitation fluence dependence of photo-excited carriers as a result of transitions through one-photon absorption (left, linear fluence dependence) or through two-photon absorption (right, quadratic fluence dependence). (b) The pump-induced change in the reflected THz electric field ($\Delta E$, orange curve) exhibits a phase difference with respect to the equilibrium reflected THz probe pulse ($E_{\text{THz}}$, grey curve). In this case, the frequency-resolved optical response is required to characterize the fluence dependence of photo-excited carriers. (c) Frequency-resolved transient change in optical conductivity, $\Delta\sigma=\sigma^{\text{transient}}-\sigma^{\text{equilibrium}}$, at the time delay of 1 ps. A Drude-like behavior is seen, indicating an influx of photo-excited carriers. Measurements for (b) and (c) taken at $T = 295$ K with 1.476 eV excitation energy.}{\label{fig:bg}}
\end{figure*}

Photo-excitation can provide critical insights into the band structure of semiconductors and insulators. The pump fluence dependence of the photo-excited carriers disentangle the underlying mechanisms through which the carriers are generated, i.e., whether the carriers originate from single-photon absorption to unoccupied states, or they come from two-photon absorption to higher energy levels (Fig.\ref{fig:bg}(a)). 
The pump wavelength dependence of photo-excited carriers can more conclusively determine the onset of interband transitions.

Here, we choose three excitation energies to drive bulk NiPS$_3$ single crystals out of equilibrium at 7 K: 1.476 eV (resonant with the ultra-narrow exciton), 1.494 eV (resonant with the magnon sideband of the exciton), and 1.610 eV (close to the optical bandgap) \cite{Kang2020, Wang2021}. In comparison, we also conduct measurements at room temperature with the same excitation energies, where the spin-orbit exciton features are no longer present \cite{Kang2020,Wang2021,Hwangbo2021}. We use time-domain terahertz (THz) spectroscopy to measure photo-excited carriers as a function of temperature, excitation fluence, and excitation energy. We evaluate the transient spectral weight of optical conductivity, which is proportional to the photo-excited carrier density. At low temperature (7K), the pump fluence dependence of the carrier density is purely quadratic for all pump energies, which we attribute to the two-photon absorption process. The absence of a linear pump fluence dependence suggests that the bandgap is above 1.610 eV at 7 K. This observation sets the lower limit of the exciton dissociation energy as 134 meV. At room temperature, the quadratic pump fluence dependence persists, while a linear fluence dependence sets in when pumping with 1.494 eV, indicating that this excitation energy is above the absorption onset. These observations provide key insights to the exciton band structure for NiPS$_3$.  

Broadband THz pulses were generated by laser-ionized plasma from a 35 fs Ti-Sapphire laser. The reflected THz signal from a bulk NiPS$_3$ single crystal was detected by electrooptical sampling in a 1 mm thick ZnTe crystal. The pump photon energies were selected by bandpass filters (10 nm bandwidth) from the same Ti-Sapphire laser used for the THz probe \cite{Warsh2024}. The incident fluence was varied from 0.45 mJ/cm$^2$ to 3.20 mJ/cm$^2$. The THz transient and the pump-induced change were simultaneously measured by two lock-in amplifiers.

A typical set of THz time-domain data is shown in figure \ref{fig:bg}(b). The light-induced change in the THz field ($\Delta E=E_{\text{THz}}^{\text{pump on}}-E_{\text{THz}}^{\text{pump off}}$) peaks at a slightly shifted position in comparison with the incident THz field ($E_{\text{THz}}=E_{\text{THz}}^{\text{pump off}}$). This data is taken at $T = 295$ K with an excitation photon energy of 1.476 eV. Similar time-domain responses are seen for all three excitation energies and for both 7 K and 295 K. The $\Delta E$ peak is sometimes referred to as spectrally-integrated transient THz response \cite{Belvin2021}, which gives an overview of the pump-induced dynamics. However, when the pump induces changes in both the amplitude and phase of the probe THz field, this frequency-integrated THz response does not correspond to a clear and meaningful quantity, such as the number of light-induced carriers. Instead, one needs to analyze the frequency-resolved optical response to characterize photo-excited carriers.

\begin{figure*}
\includegraphics[scale=1]{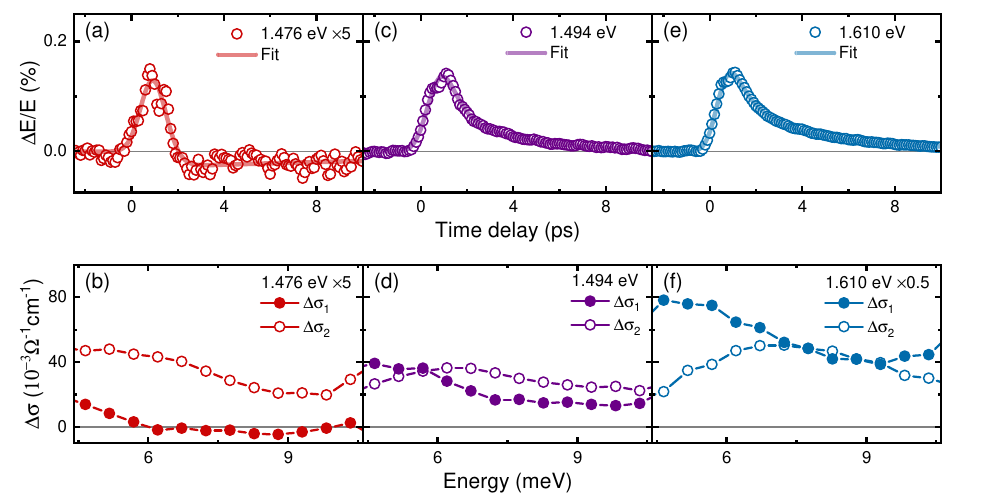}
\caption{Transient optical response at $T = 7$ K with three pumping energies at 1.476 eV, 1.494 eV, and 1.610 eV. (a, c, e) Time-domain response of the reflected THz electric field. (b, d, f) Frequency-resolved complex optical conductivity at $t = 1$ ps. The optical response at all pump energies exhibits a Drude-like spectrum.}{\label{fig:7Kresponse}}
\end{figure*}

A standard thin-film model \cite{Warsh2024} is used to evaluate the frequency-resolved optical conductivities. At maximum response (time delay $t = 1$ ps), the pump-induced transient change in optical conductivity exhibits a Drude-like shape (Fig. \ref{fig:bg}(c)). This response is indicative of an increase in free carriers in the material as a result of photo-excitation. Analysis of the fluence dependence of this carrier response allows us to unravel the mechanism through which these carriers were generated. Linear absorption of a photon with energy $\omega$, leading to the promotion of an electron to an excited state at an energy $\omega$ above its initial state, exhibits a linear fluence dependence (Fig. \ref{fig:bg}(a), left) \cite{Fox2006-Radiative, Dressel2002}. Alternatively, the material could undergo two-photon absorption (TPA), promoting an electron to an excited state at an energy $2\omega$ above its initial state. The probability of a TPA transition occurring is equivalent to the product of the probability of one photon and the probability of another photon, both of energy $\omega$, being absorbed at the same time. This results in a transition rate proportional to the square of the intensity, or an excited carrier response with quadratic fluence dependence (Fig. \ref{fig:bg}(a), right) \cite{Fox2006-Radiative}.

Before moving on to detailed analysis, we would like to point out a special feature at low temperatures. The transient optical response at $T = 7$ K exhibits a unique behavior when pumping at 1.476 eV \cite{Warsh2024}, the energy of the lowest ultra-narrow exciton peak in NiPS$_3$ \cite{Kang2020, Wang2021, Belvin2021}. For the time-domain response, the normalized reflected THz field $\Delta E/E$ shows a short increase and then a long-lived suppression, as shown in Fig. \ref{fig:7Kresponse}(a). Correspondingly, the frequency-resolved optical conductivities exhibit a narrow Drude peak and an onset of broad suppression in $\Delta\sigma_1$ (Fig. \ref{fig:7Kresponse}(b)) at $t = 1$ ps. The broad suppression evolves further and gains spectral weight with increasing time delay. It is hypothesized to be a long-lived population inversion seen at THz energies \cite{Warsh2024}. 

In contrast, for 1.494-eV and 1.610-eV pumping, $\Delta E/E$ exhibits only a positive signal with an exponential relaxation (Fig. \ref{fig:7Kresponse}(c) and (e)), as is typically seen for the case of photo-excited carriers. Correspondingly, the pump-induced transient change in optical conductivity at $t = 1$ ps for these excitation energies exhibits a simple Drude-like response (Fig. \ref{fig:7Kresponse}(d) and (f)), similar to the room-temperature response at 1.476-eV pumping as shown in Fig. \ref{fig:bg}(c). Note that in the antiferromagnetic state, the transient response should contain a magnon feature at 5.5 meV \cite{Belvin2021}. This feature is absent here due to the insufficient energy resolution, which is a result of the sample thickness (400 $\mu m$). Therefore, we only study larger energy scale features such as the bandgap by analyzing the pump fluence dependence of the photo-excited carriers. The similarity of the photo-excited carrier response, despite the resonant nature of the 1.494-eV pumping at $T$ = 7 K, indicates similar carrier generation mechanisms at these energies in both the paramagnetic and antiferromagnetic phases. In the following discussion, we will focus on the $t =1$ ps response when the Drude response is strongest for all excitation energies at all temperatures.

\begin{figure*}
\includegraphics[scale=1]{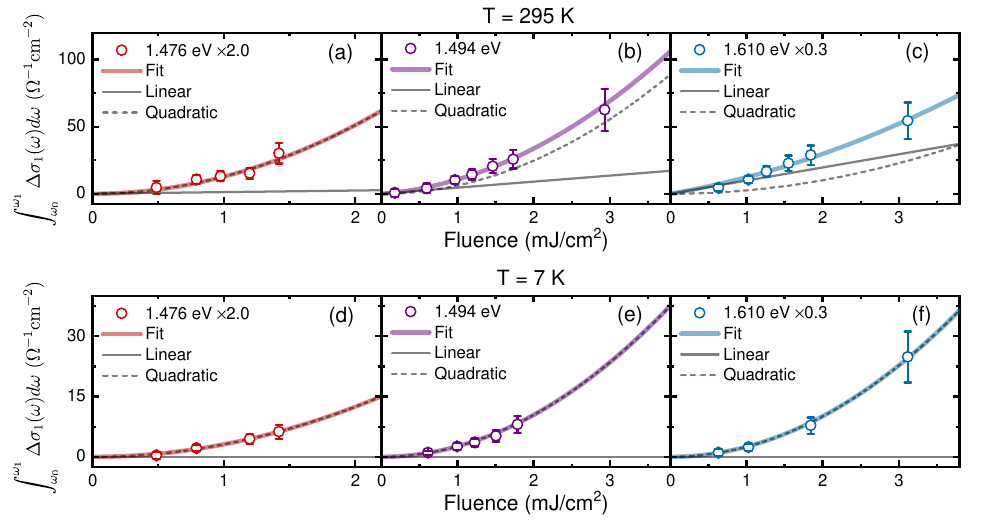}
\caption{Incident fluence dependence of the light-induced change in the partial spectral weight, $\int_{\omega_0}^{\omega_1} \Delta \sigma_1(\omega)d\omega$ in the energy region $\omega_0$ = 3.8 meV to $\omega_1$ = 11.4 meV at a fixed time delay $t = 1$ ps for three excitation energies. The transient partial spectral weight change is fit by linear and quadratic dependence components. (a-c) At $T = 295$ K, the response at all pump energies show a linear component, dominating at 1.610 eV. (d-f) At $T = 7$ K, the linear component vanishes and the quadratic component dominates at all pump energies.}{\label{fig:fd}} 
\end{figure*}

We investigate the origin of photo-excited carriers by studying the pump fluence dependence of the carrier density, which is proportional to the transient change in the partial spectral weight of the optical conductivity, defined as $\int_{\omega_0}^{\omega_1} \Delta \sigma_1(\omega)d\omega$, where $\omega_0 = 3.8$ meV and $\omega_1 = 11.4$ meV. Within this region of integration, the optical conductivity is predominantly composed of the Drude contribution, and the transient change in spectral weight is proportional to the excited free-carrier population following the optical sum rule \cite{Dressel2002}. The change in the partial spectral weight is fit with linear and quadratic fluence dependence components, as shown in Fig. \ref{fig:fd}. Although the free-electron Drude model is over simplified \cite{Ulbricht2011}, the optical sum rule is still valid for strongly correlated materials \cite{Basov2011}. Furthermore, here we study the itinerant carriers created by above-gap photo-excitation with low pump fluences. The transient optical conductivity change can be well captured by the Drude fit (see Fig. \ref{fig:bg} (c)). Similar cases in strongly correlated materials include cuprate superconductors \cite{Averitt2001,Kaindl2005}, iron-based superconductors \cite{Warshauer2023}, and antiferromagnets with unconventional excitons \cite{Belvin2021,Mehio2023,Mehio2025}, in which the Drude model and/or the optical sum rule has been applied to analyze photo-excited carriers.

\begin{figure}
\includegraphics[scale=1]{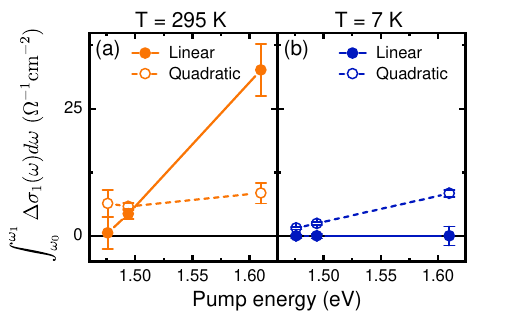}
\caption{Linear and quadratic components of the transient change in the partial spectral weight at excitation fluence 1 mJ/cm$^2$. (a) At $T = 295$ K, both linear and quadratic components exist. (b) At $T = 7$ K, no linear fluence dependence is seen.}{\label{fig:rat}}
\end{figure}

For the $T = 295$ K case, both the linear component and quadratic components are necessary to fit the fluence dependence of the partial spectral weight (Fig. \ref{fig:fd}(a-c)). The linear component is more significant in the 1.610 eV pumping case (Fig. \ref{fig:fd}(c)). At $T = 7$ K, the transient change in the partial spectral weight carries a purely quadratic dependence on fluence for all excitation energies (Fig. \ref{fig:fd}(d-f)). The linear and quadratic components of the transient partial spectral weight change at a fixed excitation fluence are illustrated in Fig. \ref{fig:rat}.

We now discuss the implications of the fluence dependence data.  The linear fluence dependence of excited free carriers, seen in the nonequilibrium response at $T = 295$ K, is characteristic of carriers generated through optical transitions from below the charge-transfer gap to the conducting states above the charge-transfer gap \cite{Dressel2002, Kittel2004-SC, Ashcroft1976-HS}. The linear component weakens with decreasing photon energy, and becomes smaller than the measurement errorbars in the 1.476 eV pumping case. Therefore, the bandgap at $T = 295$ K is smaller than and close to 1.494 eV.

For $T=7$ K, no linear fluence dependence is seen for all pump energies, indicating that the bandgap at this temperature is greater than 1.610 eV. Insulating or semiconducting materials commonly exhibit temperature dependent energy gaps. The primary causes of this temperature dependence are thermal expansion of the lattice's periodic potential and increased carrier-phonon interaction with increasing temperature \cite{Ashcroft1976-HS, Unlu1992, Mann2024, Bhosale2012, Kittel2004-SC}. A temperature dependent bandgap is also in agreement with earlier characterization of excitonic features: the spin-orbit-entangled exciton peaks of NiPS$_3$ shift to higher energies with decreasing temperature below the N\'eel temperature $T_N = 155$ K \cite{Kang2020, Wang2021, Hwangbo2021, Ho2021}. 
The spin-orbit entangled exciton transition involves an exchange between the S $3p$ orbitals and partially filled Ni $3d$ orbitals which comprise the charge-transfer gap in NiPS$_3$ \cite{Kang2020, He2024, Ho2021}. 
As the bandgap is larger than 1.610 eV at 7 K, and the exciton feature peaks at 1.476 eV, we estimate the lower limit of the exciton dissociation energy to be 134 meV, i.e. the energy difference between the spin-correlated exciton and the corresponding charge-transfer continuum \cite{Belvin2021, Monney2016}.
The absorption onset blueshifts about 100 meV from room temperature to 10 K \cite{Ho2021}, which agrees well with the energy difference of the bandgap from our estimation (close to 1.494 eV at room temperature and greater than 1.610 eV at 7 K). Therefore, we expect that the bandgap at 7 K is close to 1.610 eV, and 134 meV is close to the exact value of exciton dissociation energy for bulk NiPS$_3$.

Finally, we discuss the origin of the quadratically excited carriers. In contrast to the large difference in linearly excited carriers, all pump energies at $T = 295$ K exhibit similar quadratically excited spectral weight, as shown in Fig. \ref{fig:rat}. Strong absorption and conductivity at energies between 2.95 eV and 3.22 eV give validity to TPA as a driving mechanism for carrier generation \cite{Kim2018}. In addition, angle-resolved photoemission spectroscopy measurements of NiPS$_3$ show unoccupied bands with significant density of states available at energies above 2.3 eV \cite{Cao2024}. The similar, large carrier densities between all pump energies at T = 295 K would indicate that the quadratically excited carriers were all generated through the same TPA process which accesses energy levels well above the energy of first conducting states. The reduction of the quadratic component for lower pump energies at T = 7 K is due to the strong excitonic absorption features at 1.476 eV and 1.494 eV below the N\'eel temperature. Additionally, the narrowing of the 3.5 eV absorption peak with decreasing temperature results in fewer TPA states for the lower transition energies \cite{Kim2018}. A quadratic fluence dependence of the photo-excited carriers can also arise from exciton-exciton annihilation \cite{Braun1968, Sun2014}. Note that the exciton feature disappears close to the Ne\'el temperature \cite{Kang2020,Wang2021,Hwangbo2021}, therefore the exciton-exciton annihilation mechanism is not relevant for the quadratic fluence dependence at room temperature. Such a mechanism could be at play for the 7 K case with 1.476-eV pumping. However, there is no significant difference in the quadratic partial spectral weight at 7 K versus at 295 K, as shown in Fig. \ref{fig:rat}. In addition, the lifetime of the photo-excited carrier contribution to the change in spectral weight does not change with temperature \cite{Warsh2024}. Therefore, the photo-excited carriers which carry a quadratic fluence dependence likely share the same origin in the resonant and off-resonant cases at 7 K and 295 K. Finally, a different origin of the quadratic fluence dependence does not alter our conclusions on the bandgap, estimated from the linear pump fluence dependence data.

In conclusion, we use the pump fluence and wavelength dependence of the photo-excited carrier density to identify the bandgap values and the exciton dissociation energy for NiPS$_3$. The bandgap is close to 1.494 eV at room temperature and near 1.610 eV at 7K. As the exciton feature is seen at 1.476 eV, we estimate the exciton dissociation energy to be close to 134 meV. These results are vital to quantify the  characteristics of the unconventional spin-correlated exciton and to validate band structure calculations for NiPS$_3$, which is a unique candidate for future two-dimensional spintronic applications.

\begin{acknowledgments}
This material is based upon work supported by the National Science Foundation under Grant No. 1944957. H. C. and W. H. acknowledge support from the U.S. Department of Energy, Office of Science, Office of Basic Energy Sciences Early Career Research Program under Award Number DE-SC-0021305. Work by Q. T., J. T. and X. L. were supported by the National Science Foundation under Grant No. 1945364 and the U.S. Department of Energy, Office of Science, Basic Energy Sciences under Award DE-SC0021064. Q.T. acknowledges support of the Laursen Graduate Research Award. We acknowledge the helpful discussions with Matthias Hoffmann,  Zhuquan Zhang and Sahar Sharifzadeh. We thank Boston University Photonics Center for technical support.
\end{acknowledgments}

\textbf{Data Availability.} The supporting data for this article are openly available on Harvard Dataverse \cite{WarshData}.


\end{document}